\documentclass[twocolumn,aps,showpacs,preprintnumbers,amsmath,amssymb, superscriptaddress]{revtex4-1}
\pdfoutput=1
\usepackage{bm}
\usepackage{graphicx}
\usepackage{color}
\usepackage{xspace}
\definecolor{dblue}{rgb}{0,0,0.6}
\definecolor{dred}{rgb}{0.9,0,0}
\definecolor{dgreen}{rgb}{0,0.4,0}
\usepackage{ulem}

\begin{document}

\title{Collective excitations of dynamic Fermi surface deformations in BaFe$_2$(As$_{0.5}$P$_{0.5}$)$_2$}  
   
\author{S.-F. Wu}
\affiliation{Department of Physics and Astronomy, Rutgers University, Piscataway, NJ 08854, USA}
\affiliation{Beijing National Laboratory for Condensed Matter Physics, and Institute of Physics, Chinese Academy of Sciences, Beijing 100190, China}
\author{W.-L. Zhang}
\affiliation{Beijing National Laboratory for Condensed Matter Physics, and Institute of Physics, Chinese Academy of Sciences, Beijing 100190, China}
\author{D. Hu}
\affiliation{Beijing National Laboratory for Condensed Matter Physics, and Institute of Physics, Chinese Academy of Sciences, Beijing 100190, China}
\author{H.-H. Kung}
\affiliation{Department of Physics and Astronomy, Rutgers University, Piscataway, NJ 08854, USA}
\author{A. Lee}
\affiliation{Department of Physics and Astronomy, Rutgers University, Piscataway, NJ 08854, USA}
\author{H.-C.~Mao}
\affiliation{Beijing National Laboratory for Condensed Matter Physics, and Institute of Physics, Chinese Academy of Sciences, Beijing 100190, China}
\author{P.-C. Dai}
\affiliation{Department of Physics and Astronomy, Rice University, Houston, Texas 77005, USA}
\author{H. Ding}
\affiliation{Beijing National Laboratory for Condensed Matter Physics, and Institute of Physics, Chinese Academy of Sciences, Beijing 100190, China}
\affiliation{Collaborative Innovation Center of Quantum Matter, Beijing, China}
\author{P. Richard}\email{p.richard@iphy.ac.cn}
\affiliation{Beijing National Laboratory for Condensed Matter Physics, and Institute of Physics, Chinese Academy of Sciences, Beijing 100190, China}
\affiliation{Collaborative Innovation Center of Quantum Matter, Beijing, China}
\author{G. Blumberg}\email{girsh@physics.rutgers.edu}
\affiliation{Department of Physics and Astronomy, Rutgers University, Piscataway, NJ 08854, USA}
\affiliation{National Institute of Chemical Physics and Biophysics, 12618 Tallinn, Estonia}
\date{\today}
 
\begin{abstract}
We use electronic Raman scattering to study the low-energy excitations in BaFe$_2$(As$_{0.5}$P$_{0.5}$)$_2$ ($T_c \approx 16$ K) samples. In addition to a superconducting pair breaking peak (2$\Delta=6.7$ meV) in the A$_{1g}$ channel with a linear tail towards zero energy, suggesting a nodal gap structure, we detect spectral features associated to Pomeranchuk oscillations in the A$_{1g}$, B$_{1g}$ and B$_{2g}$ channels. We argue that the small Fermi energy of the system is an essential condition for these Pomeranchuk oscillations to be underdamped. The Pomeranchuk oscillations have the same frequencies in the B$_{1g}$ and B$_{2g}$ channels, which we explain by the mixing of these symmetries resulting from the removal of the $\sigma_v$ and $\sigma_v$ symmetry planes due to a large As/P disorder. Interestingly, we show that the temperature at which the peaks corresponding to the Pomeranchuk oscillations get underdamped is consistent with the non-Fermi liquid to Femi liquid crossover determined by transport, suggesting that the Pomeranchuk instability plays an important role in the low-energy physics of the Fe-based superconductors. 
\end{abstract} 


\maketitle

With their multiband nature, the Fe-based superconductors provide an interesting playground for studying many-body effects and collective excitations. Of particular interest is the role played by the degeneracy of the $d_{xz}$ and $d_{yz}$ orbitals. Many Fe-based superconductors exhibit an antiferromagnetic phase transition that is preceded by a small structural distortion at a temperature $T_s$ that removes the degeneracy between the $d_{xz}$ and  $d_{yz}$ orbitals. For BaFe$_2$(As$_{1-x}$P$_x$)$_2$, evidence was shown for the existence of quantum critical behavior at $x=0.3$ \cite{Nakai_PRL105,Hashimoto_Science336,Walmsley_PRL110,Analytis2014NatPhy}, which coincides with the composition at which $T_s$ drops to zero. At that particular doping, the electrical resistivity $\rho(T)$ at high temperature varies linearly \cite{Hashimoto_Science336,Analytis2014NatPhy}, which is attributed to a non-Fermi liquid (NFL) behavior. This NFL behavior is found over a wide range of temperature and P concentration, and it is separated by a crossover from a Fermi liquid (FL) like regime with $\rho(T)\sim T^2$ \cite{Analytis2014NatPhy}. 

Mainly due to the complexity of the interplay between the spin and orbital degrees of freedom \cite{Fernandes2014NatPhy}, there is still no consensus on the precise nature of the critical fluctuations preceding the magnetic and structural transitions, which may also be responsible for the observed NFL behavior. Recently, Raman scattering studies evidenced the existence of quadrupolar charge fluctuations with the B$_{2g}$ symmetry above $T_s$ in NaFe$_{1-x}$Co$_x$As \cite{Thorsmolle2016PRB}, Ba(Fe$_{1-x}$Co$_x$)$_2$As$_2$ \cite{Gallais2013PRL,Kretzschmar_Nature12}, $A$Fe$_2$As$_2$ [$A$ = (Sr, Eu)] \cite{ZhangWL2014arxiv} and FeSe \cite{Massat_arxivFeSe}. Such Pomeranchuk like instability \cite{Pomeranchuk1958}, where the Fermi surface (FS) encounters dynamical deformation, is a natural consequence of the degeneracy of the $d_{xz}$ and $d_{yz}$ orbitals at high temperature. The corresponding susceptibility, which can be probed directly by Raman scattering, diverges as approaching the Pomeranchuk instability. However, there is to date no experimental correlation between the Pomeranchuk fluctuations \cite{Pomeranchuk1958,Oganesyan2001PRB}  and the NFL behavior observed in BaFe$_2$(As$_{1-x}$P$_{x}$)$_2$, and whether the fluctuations may appear in the other symmetry channels, is unclear. 

In this Letter, we use electronic Raman scattering to study the low-energy excitations in BaFe$_2$(As$_{0.5}$P$_{0.5}$)$_2$ ($T_c\approx 16$ K), for which there is neither long-range magnetic nor long-range nematic ordering. We reveal evidence for Pomeranchuk oscillations in different symmetry channels, with frequencies larger than the superconducting (SC) gap determined from a SC pair breaking peak (2$\Delta=6.7$ meV) in the A$_{1g}$ channel. We argue that the small Fermi energy of the system is an essential condition for these Pomeranchuk oscillations to be underdamped. No sharp collective mode is observed below the SC gap, possibly due to the nodal nature of the SC gap structure derived from the finite and linear Raman response towards zero energy in the A$_{1g}$ channel. Interestingly, the Pomeranchuk oscillations have the same frequency in the B$_{1g}$ and B$_{2g}$ channels, which we explain by the mixing of these symmetries resulting from the removal of the $\sigma_v$ and $\sigma_v$ symmetry planes due to a large As/P disorder. More importantly, we show that the temperature at which the peaks corresponding to the Pomeranchuk oscillations get underdamped is consistent with the FL to NFL crossover determined by transport. This suggests that the Pomeranchuk instability plays an essential role in shaping the electronic properties of Fe-based superconductors.

\begin{figure}[!t]
\begin{center}
\includegraphics[width=3.4in]{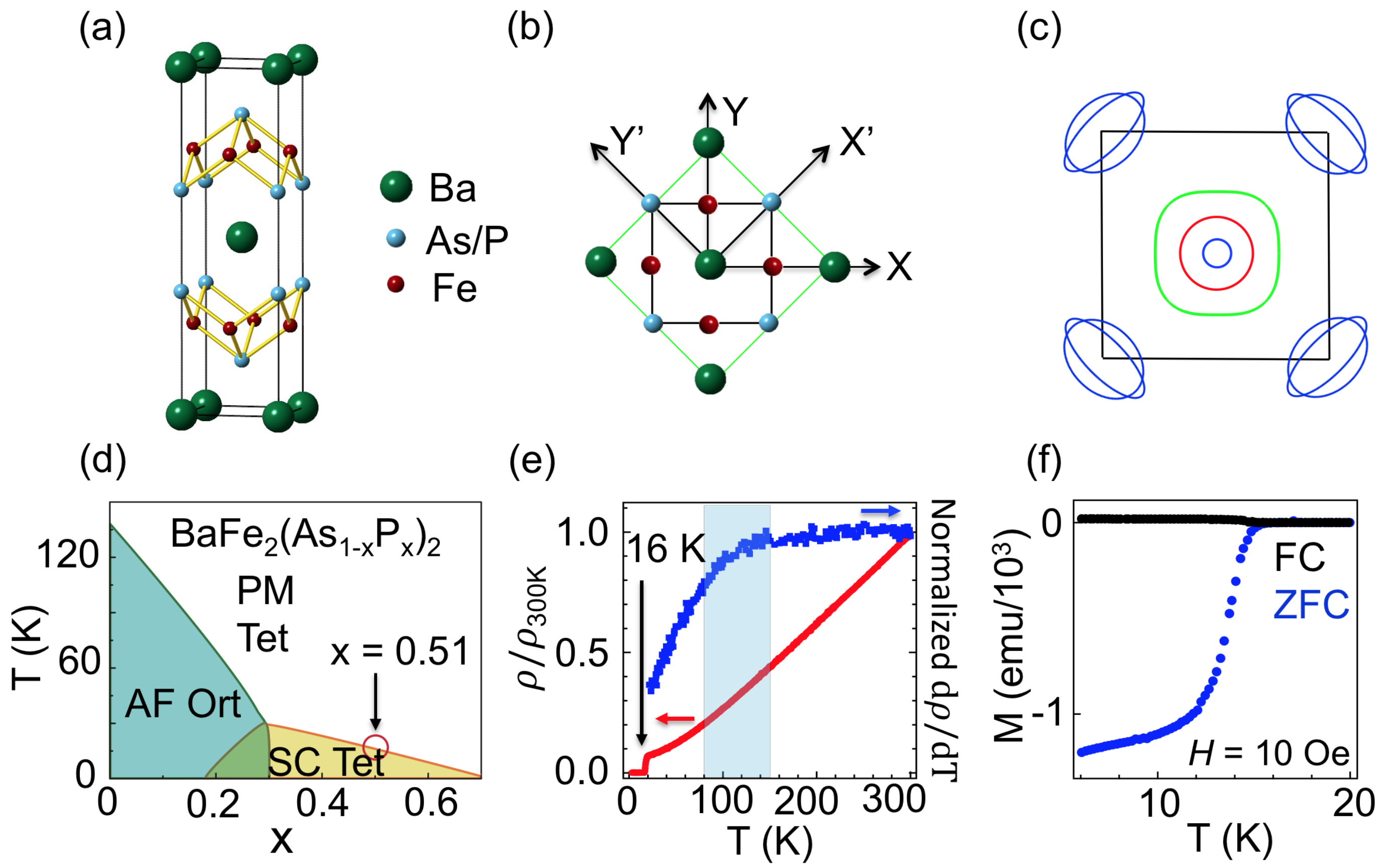}
\end{center}
\caption{\label{Fig1_transport}(Color online) Properties of BaFe$_2$(As$_{0.5}$P$_{0.5}$)$_2$. (a) Crystal structure of BaFe$_2$(As$_{0.5}$P$_{0.5}$)$_2$. (b) Definition of the X, Y, X$'$ and Y$'$ directions. The green and black lines represent the 4-Fe and 2-Fe unit cells, respectively. (c) Schematic representation of the FSs in the 2-Fe Brillouin zone in the $k_z=\pi$ plane \cite{Yoshida2011PRL,suzuki2014PRB}. (d) Schematic phase diagram of BaFe$_2$(As$_{1-x}$P$_{x}$)$_2$ \cite{Hu2015PRL}.  (e) Normalized resistivity (red) and its normalized derivative (blue). The shadow area indicates a crossover from quadratic to linear temperature dependence. (f) Magnetic susceptibility in field-cooled (FC) and zero-field-cooled (ZFC) modes.}
\end{figure}

Single crystals of BaFe$_2$(As$_{0.5}$P$_{0.5}$)$_2$ were grown using the Ba$_{2}$As$_{3}$/Ba$_{2}$P$_{3}$ self-flux method described in Ref. \cite{Masamichi2012JPSJ}, and the chemical compositions were determined by inductive coupled plasma analysis. The crystal structure, illustrated in Figs. \ref{Fig1_transport}(a) and \ref{Fig1_transport}(b), belongs to space group $I4/mmm$ (point group D$_{4h}$). Resistivity (Fig. \ref{Fig1_transport}(e)) and magnetic susceptibility (Fig. \ref{Fig1_transport}(f)) measurements indicate $T_c \approx 16$ K. The crystals used for Raman scattering were cleaved and positioned in a continuous He flow optical cryostat. The measurements presented here were performed in a quasi-back scattering geometry along the crystallographic $c$-axis using the Kr$^+$ laser line at 647.1~\,nm (1.92 eV). The excitation laser beam was focused into a $50\times100$ $\mu$m$^2$ spot on the $ab$-surface, with the incident power smaller than 10 and 6 mW for measurements in the normal state and in the SC state, respectively. The scattered light was collected and analyzed by a triple-stage Raman spectrometer designed for high-stray light rejection and throughput, and then recorded using a liquid nitrogen-cooled charge-coupled detector. Raman spectra were recorded in the range of 10 to 350~cm$^{-1}$ using a 1800 grooves/mm grating and in the range of 100 to 2000~cm$^{-1}$ using a 150 grooves/mm grating. Raman scattering intensity data were corrected for the spectral responses of the spectrometer and detector. The temperature has been corrected for laser heating. 

In this manuscript, we define X and Y along the 2-Fe unit cell basis vector (at 45$^{\circ}$ degrees from the Fe-Fe direction) in the tetragonal phase, whereas X$'$ and Y$'$ are along the Fe-Fe directions, as shown in Fig.~\ref{Fig1_transport}(b). We also define R and L as circular right and circular left polarizations. The static Raman susceptibility $\chi'(0,T)$ was obtained from a Kramers-Kronig transformation with a high-energy cut-off at 350 cm$^{-1}$. The B$_{1g}$ phonon was removed by fitting before the Kramers-Kronig transformation in the B$_{1g}$ channel.

\begin{figure}[!t]
\begin{center}
\includegraphics[width=\columnwidth]{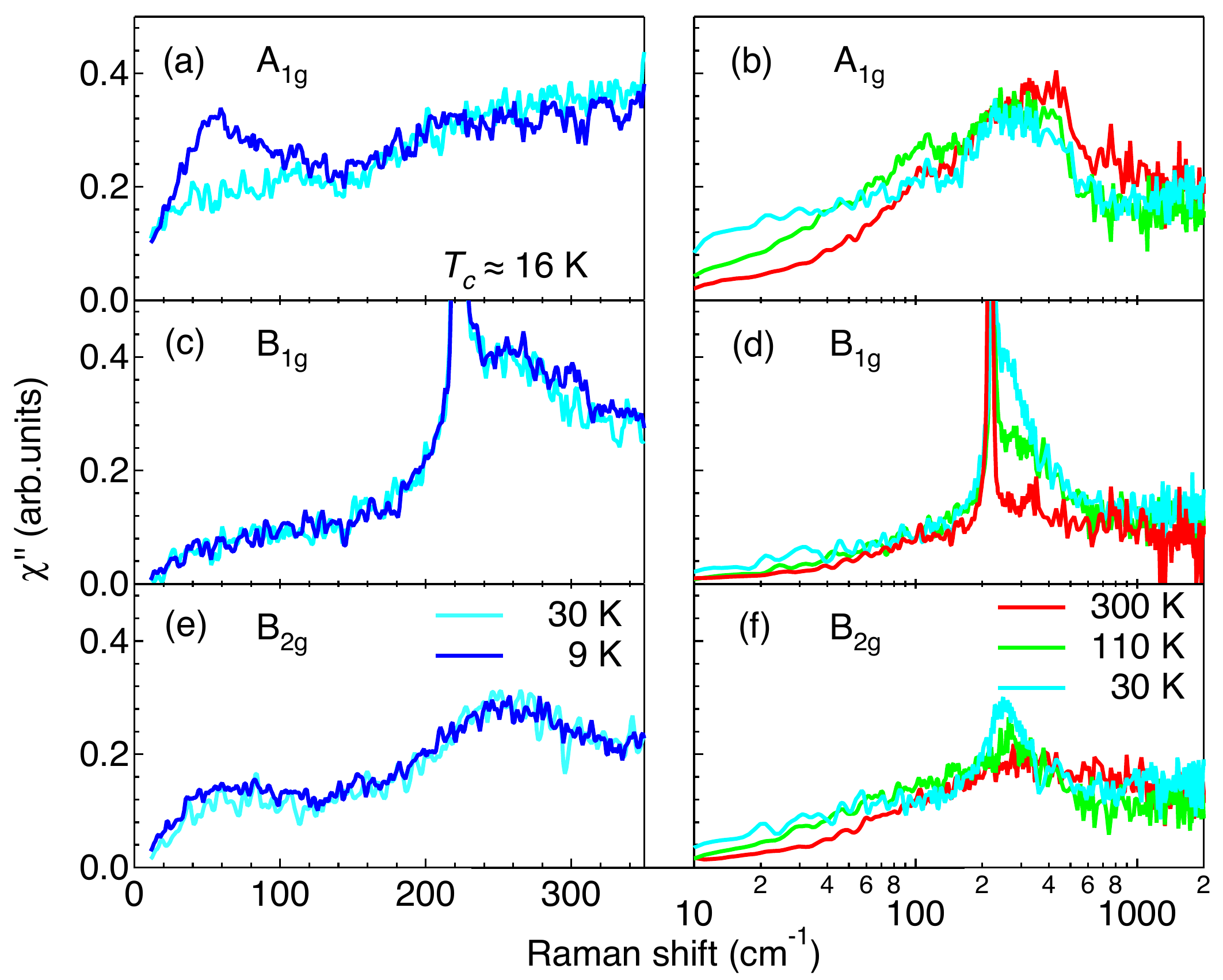}
\end{center}
\caption{\label{Fig2_A1gB1gB2g}(Color online) Raman response $\chi''_{A_{1g}}(\omega,T)$ in (a-b), $\chi''_{B_{1g}}(\omega,T)$ in (b-c) and  $\chi''_{B_{2g}}(\omega,T)$ in (c-d) of BaFe$_2$(As$_{0.5}$P$_{0.5}$)$_2$ at different temperatures.}
\end{figure}

The Raman selection rules for crystals with the D$_{4h}$ point group symmetry indicate that the XX, XY, X$'$X$'$, X$'$Y$'$, RR, RL polarization geometries probe the A$_{1g}$ + B$_{1g}$, A$_{2g}$+ B$_{2g}$, A$_{1g}$ + B$_{2g}$, A$_{2g}$ + B$_{1g}$, A$_{1g}$ + A$_{2g}$ and B$_{1g}$ + B$_{2g}$ channels, respectively. Assuming a same featureless luminescence background $I_{BG}$ for all symmetry channels and that the A$_{2g}$ response is negligible, the imaginary part of the Raman susceptibility $\chi''$ in the different symmetry channels can easily be isolated.

In Figs. \ref{Fig2_A1gB1gB2g}(a)-\ref{Fig2_A1gB1gB2g}(b), we show the Raman susceptibility $\chi''_{A_{1g}}(\omega,T)$ at different temperatures from 300 K to 9 K. The most prominent feature that appears in the low-energy part of the A$_{1g}$ channel (Fig. \ref{Fig2_A1gB1gB2g}(a)) is a peak emerging at 54 cm$^{-1}$ (6.7 meV) at 9 K, in the SC state. Tentatively ascribing this feature to a pair breaking peak would lead to $2\Delta=6.7$ meV. Such value is about twice lower than the values of SC gaps reported by ARPES for the optimal-$T_c$ compound \cite{zhang2012NatPhy}, for which $T_c$ is about twice larger. Another possible interpretation for this peak, also related to superconductivity, is an electron-hole plasmonic excitation inside the SC gap, as reported earlier in NaFe$_{1-x}$Co$_x$As \cite{Thorsmolle2016PRB}. Unlike the expectation for a completely open gap, the spectral intensity decreases linearly towards the low frequencies, indicating the presence of excitations with energy lower than $2\Delta$. This suggests that the SC gap probed in the A$_{1g}$ channel is nodal. Although the detail of the SC gap structure measured by ARPES on BaFe$_2$(As$_{1-x}$P$_{x}$)$_2$ is still controversial \cite{shimojima2011orbital,zhang2012NatPhy,yoshida2014SciRep}, strong SC gap anisotropy \cite{yoshida2014SciRep} and even an horizontal line node \cite{zhang2012NatPhy} have been proposed. A nodal gap structure is also inferred from thermal conductivity measurements \cite{YamashitaPRB84,Qiu_PRX2}.

In Fig. \ref{Fig2_A1gB1gB2g}(b), we show the A$_{1g}$ Raman spectra up to 2000 cm$^{-1}$ in a semi-log plot. We detect a broad peak centered around 350 cm$^{-1}$, with a full-width-at-half-maximum (FWHM) of about 300 cm$^{-1}$. Upon cooling, this peak shows some temperature dependence, but the most obvious changes occur in the low-energy part of the spectrum, which is enhanced at low temperature.
 
\begin{figure}[t] 
\begin{center}
\includegraphics[width=3.4in]{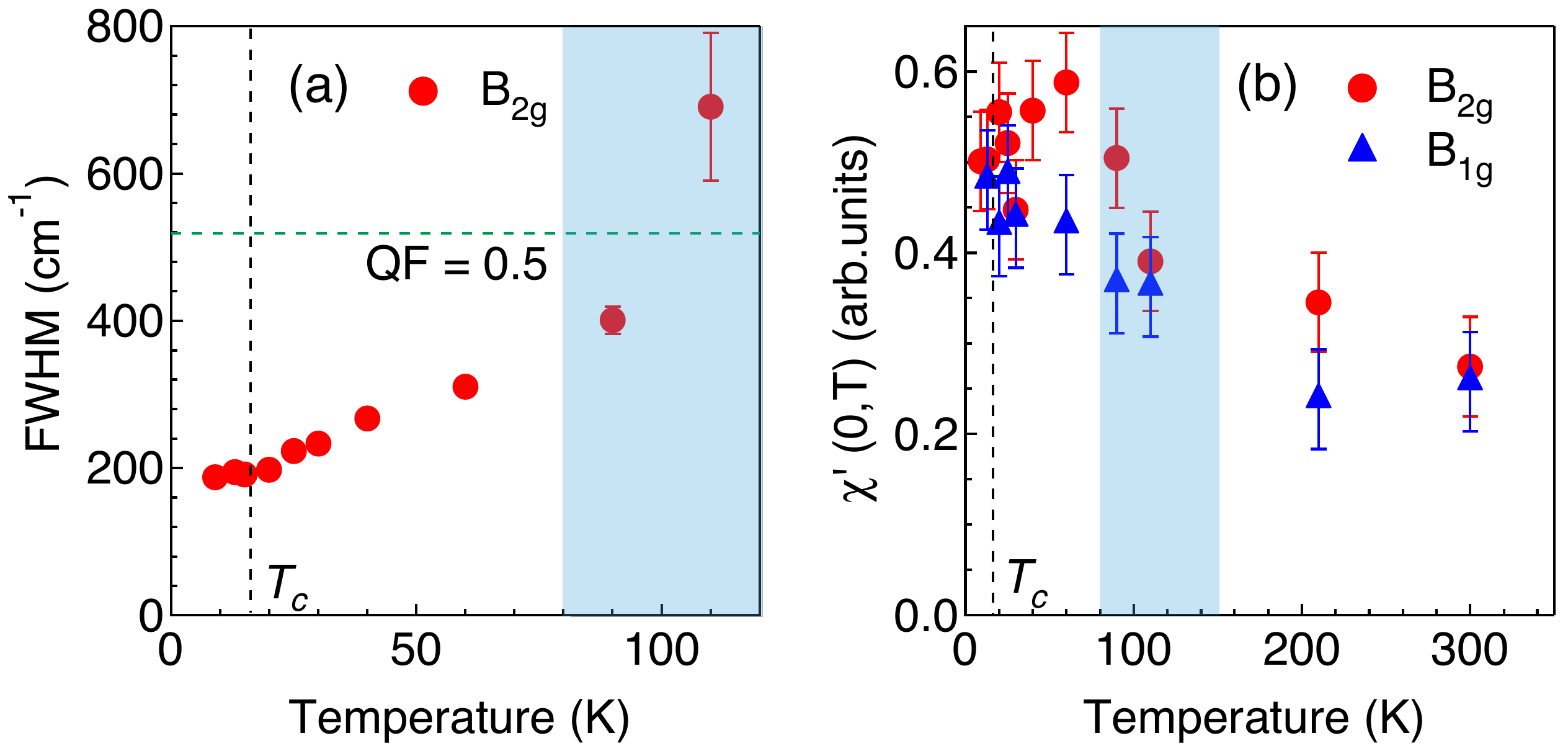}
\end{center}
\caption{\label{Fig3_Chi0}(Color online) (a) Static Raman susceptibility $\chi'(0,T)$ in the B$_{1g}$ (blue triangles) and B$_{2g}$ (red circles) channels. The green horizontal dashed line corresponds to critical damping (QF = 1/2). (b) FWHM of the 260~cm$^{-1}$ peak in the B$_{2g}$ channel. The vertical dashed lines and the blue shadow areas in (a) and (b) indicate $T_c$ and the crossover from $T^2$ to $T$ resistivity, respectively. }
\end{figure}

The $\chi''_{B_{1g}}(\omega,T)$ susceptibility displayed in Figs. \ref{Fig2_A1gB1gB2g}(c)-\ref{Fig2_A1gB1gB2g}(d) shows a sharp peak at 218 cm$^{-1}$ (300 K) that is ascribed to a B$_{1g}$ phonon \cite{Rahlenbeck2009PRB}. It corresponds to the vibration of Fe atoms along the $c$ axis. A peak at 260~cm$^{-1}$ develops upon cooling on the right side of the phonon mode. This peak is asymmetric, with a high-energy tail extending to 800 cm$^{-1}$, as shown in Fig. \ref{Fig2_A1gB1gB2g}(d). The peak reaches half of its highest intensity at 110 K, and its maximum intensity saturates at low temperature, with no obvious change across $T_c$ (see Fig. \ref{Fig2_A1gB1gB2g}(c)). The temperature dependence of the FWHM, displayed in Fig. \ref{Fig3_Chi0}(a), is nearly 200 cm$^{-1}$ at 9 K, which leads to a quality factor (QF) of 1.3 (underdamped regime). In contrast, QF smaller than 0.5 (overdamped regime) is observed for a FHWM larger than 520 cm$^{-1}$, which occurs slightly above 100 K.

A recent study of the resistivity across the phase diagram of BaFe$_2$(As$_{1-x}$P$_{x}$)$_2$ revealed a crossover between a regime with $\rho(T)\sim T$ at high temperature and a regime $\rho(T)\sim T^2$ at low temperature \cite{Analytis2014NatPhy}. As shown by the first derivative of the resistivity in Fig. \ref{Fig1_transport}(e), this is also what we find in our samples, with a crossover regime spanning from about 80 to 150 K. As indicated by the shadow area in Fig. \ref{Fig3_Chi0}(a), this temperature regime coincides with the temperature at which the Raman peak at 260~cm$^{-1}$ switches from overdamped to underdamped (QF = 1/2), suggesting that Raman scattering probes the same crossover as measured in transport.

Interestingly, the $\chi''_{B_{2g}}(\omega,T)$ susceptibility displayed in Figs. \ref{Fig2_A1gB1gB2g}(e)-\ref{Fig2_A1gB1gB2g}(f) also shows a peak at 260~cm$^{-1}$ that is absent at 300 K but gets enhanced upon cooling. The broadness of the peak and the temperature at which it appears are very similar to what is observed for the 260~cm$^{-1}$ mode detected in the B$_{1g}$ channel, suggesting that these two peaks have the same origin. Although such degeneracy is not expected within the $D_{4h}$ point group symmetry, one can argue that the randomness of the As-P distribution near $x=0.5$ locally breaks the $\sigma_v$ and $\sigma_d$ symmetry planes, allowing mixing the B$_{1g}$ and B$_{2g}$ symmetries. In fact, the absence in our data of the A$_{1g}$ pnictogen phonon mode observed in other Fe-based superconductors \cite{Chauvilere_PRB80,Rahlenbeck2009PRB,Um_PRB85,Thorsmolle2016PRB}, is a good indication of the disorder effect. We caution that since As and P are isovalent, the removal of the $\sigma_v$ and $\sigma_d$ symmetry planes is a relatively weak perturbation that would not affect modes that do not involve the pnictogen atoms, such as the Fe-related B$_{1g}$ phonon, which remains sharp and for which no leak is detected in the B$_{2g}$ channel. In addition to the peak at 260~cm$^{-1}$, we observe in the B$_{2g}$ channel an extra peak at 60 cm$^{-1}$ that shows enhancement in the SC state.

We show the temperature dependence of $\chi'(0,T)$ in Fig. \ref{Fig3_Chi0}(b). The $\chi'(0,T)$ obtained in both channels is enhanced gradually upon cooling from 300 K to around 50~K, \textit{i. e.} above $T_c$, before being gradually weakened at lower temperature. The increase of $\chi'(0,T)$ is possibly due to the low-frequency relaxation accros the NFL to FL crossover indicated by a shadow area in Fig. \ref{Fig3_Chi0}(b).  
   
\begin{figure}[!t] 
\begin{center} \
\includegraphics[width=3.4in]{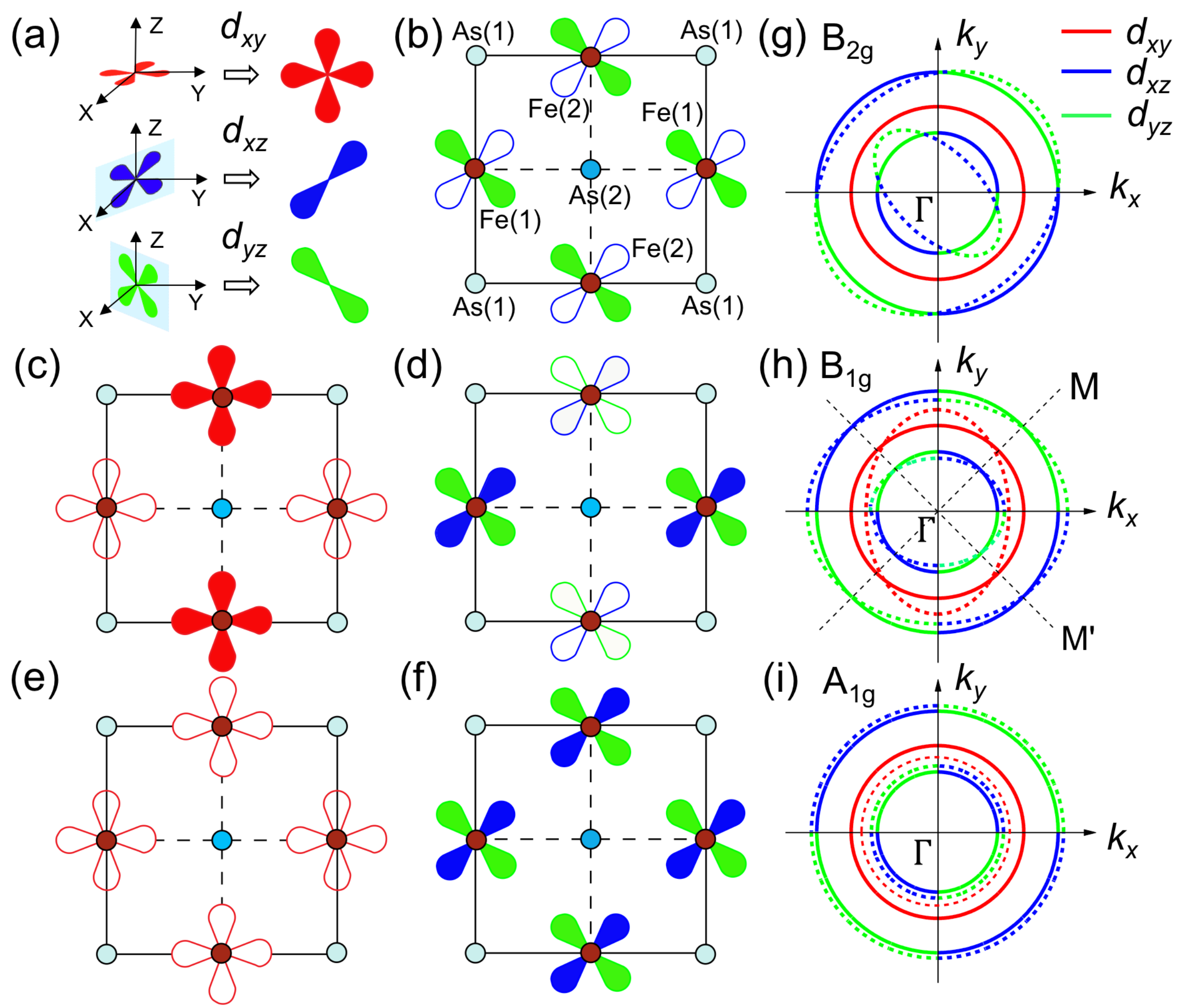}
\end{center}
\caption{\label{Fig4_FS_distortion}(Color online) Cartoons of symmetrized Pomeranchuk oscillations inn real and momentum spaces. (a) Real space orientation of the $d_{xy}$ and $d_{xz}/d_{yz}$ orbitals. (b) B$_{2g}$ type charge transfer within the $d_{xz}/d_{yz}$ orbitals. Adjacent Fe sites are labeled Fe(1) and Fe(2).  As sites above and below the Fe plane are labeled As(1) and As(2). (c-d) shows B$_{1g}$ type anti-phase charge transfer from $d_{xy}$ to $d_{xz}/d_{yz}$ orbitals on one Fe site and the opposite direction on a neighbor site. (e-f) A$_{1g}$ type of in-phase charge transfer from $d_{xy}$ to $d_{xz}/d_{yz}$ orbitals on all Fe sites. FS deformations of (g) B$_{2g}$ type, (h) B$_{1g}$ type, and (i) A$_{1g}$ type.}
\end{figure}

We now try to provide a physical picture for the mode at 260~cm$^{-1}$. In recent Raman scattering studies the presence of a low-energy excitation with the B$_{2g}$ symmetry is interpreted in terms of quadrupolar charge fluctuations [orbital singlet excitations ($\Delta L=2$)] or Pomeranchuk instability \cite{Thorsmolle2016PRB,Kretzschmar_Nature12,ZhangWL2014arxiv,Massat_arxivFeSe}. Indeed, a dynamic charge quadrupole moment of B$_{2g}$ symmetry  (nodes along the X-Y directions) can form locally on Fe sites due to the fluctuating partial charge transfer between degenerate $d_{xz}$ and $d_{yz}$ orbitals in the tetragonal phase \cite{Thorsmolle2016PRB,ZhangWL2014arxiv}. In Fig. \ref{Fig4_FS_distortion}(b), we show a cartoon of B$_{2g}$ symmetry charge transfer oscillation from $d_{xz}$ to $d_{yz}$. In momentum space, the charge transfer leads to dynamical FS distortions of the $d_{xz}/d_{yz} $ bands with nodes along $\Gamma$-X and $\Gamma$-Y, as illustrated in Fig. \ref{Fig4_FS_distortion}(g). In NaFe$_{1-x}$Co$_x$As, the peak corresponding to the quadrupolar charge fluctuations is found at low energy \cite{Thorsmolle2016PRB}. Above $T_c$, the excitation is broad and it gets completely overdamped at high temperature. However, below $T_c$ the mode becomes extremely sharp due to the absence of low-energy particle-hole excitations following a full SC gap opening \cite{ZH_Liu_PRB84,Thirupathaiah_PRB86,QQ_Ge_PRX3}.

Because the FS topology of BaFe$_2$(As$_{0.5}$P$_{0.5}$)$_2$ is similar to that of NaFe$_{1-x}$Co$_x$As, one should also expect to observe Pomeranchuk oscillations in BaFe$_2$(As$_{0.5}$P$_{0.5}$)$_2$, with the B$_{2g}$ symmetry. For this reason, it is natural to attribute the origin of the mode at 260~cm$^{-1}$ to charge quadrupole fluctuations. However, we caution that the physical situation may be different in BaFe$_2$(As$_{0.5}$P$_{0.5}$)$_2$. For example, the SC gap in BaFe$_2$(As$_{0.5}$P$_{0.5}$)$_2$ may be nodal, as discussed above, and thus particle-hole excitations can always damp the Pomeranchuk oscillations, thus preventing the emergence of a very sharp mode at low energy, in the SC phase. With a QF of 1.3, the mode at 260~cm$^{-1}$, above the SC gap, is quite sharp. This underdamped mode is possible only if the Pomeranchuk oscillation frequency $\omega_B$ is large compared to thermal fluctuations and to the Fermi energy of the system. The first condition, $T<<\omega_B$, is respected since 260~cm$^{-1}$ corresponds to a thermal energy of 374 K. Based on ARPES data on BaFe$_2$(As$_{0.5}$P$_{0.5}$)$_2$ \cite{YeZR2012PRB}, we estimate that while the holelike $d_{xy}$ band tops at about 30 meV at the $\Gamma$ point, the two $d_{xz}/d_{yz}$ bands top at 10 meV and 15~meV. This means that the Fermi energy of the system is smaller than $\omega_B$, which is consistent with underdamped Pomeranchuk oscillations (QF $>1/2$). 
 
As discussed previously, it is likely that the 260~cm$^{-1}$ excitation in the B$_{1g}$ channel is related to the Pomeranchuk oscillation found in the B$_{2g}$ channel at the same energy. We note that BaFe$_2$(As$_{0.5}$P$_{0.5}$)$_2$ is significantly more 3D than the (Ba,K)Fe$_2$As$_2$ \cite{Analytis2010PRL,Shishido2010PRL,YeZR2012PRB}. While it forms the outermost $\Gamma$-centered FS pocket in the $k_z=0$ plane, the nearly non-$k_z$-dispersive $d_{xy}$ pocket in BaFe$_2$(As$_{0.7}$P$_{0.3}$)$_2$ has a size comparable to one of the $d_{xz}/d_{yz}$ FS (the odd combination of the $d_{xz}$ and $d_{yz}$ orbitals) at $k_z=\pi$ \cite{zhang2012NatPhy}. Consequently, one can expect enhanced scattering between the $d_{xy}$ and $d_{xz}/d_{yz}$ FSs, and thus possible charge transfer. The scenario of an anti-phase $d_{xy}-d_{xz}/d_{yz}$ charge transfer is illustrated in Figs. \ref{Fig4_FS_distortion}(c) and \ref{Fig4_FS_distortion}(d). In particular, on-site $d_{xy}\rightarrow d_{xz}/d_{yz}$ charge transfer occurring in opposite directions for neighbor Fe sites leads to the formation of two charge quadrupole moments with B$_{1g}$ symmetry.  Their nodes are along the X$'$\text{-}Y$'$ directions. In momentum space, these fluctuating charge transfer can induce a dynamic FS distortions with nodes along $\Gamma$-M and $\Gamma$-M', as illustrated in Fig. \ref{Fig4_FS_distortion}(h). As mentioned above, the removal of the $\sigma_v$ and $\sigma_d$ symmetry operations due to the As/P disorder couples the B$_{1g}$ and B$_{2g}$ symmetries, providing an explanation for the same oscillation frequencies in the B$_{1g}$ and B$_{2g}$ channels. 

For completeness, we show in Figs. \ref{Fig4_FS_distortion}(e) and \ref{Fig4_FS_distortion}(f) that in-phase charge transfer between the $d_{xy}$ orbital and the $d_{xz}/d_{yz}$ orbitals is also possible. The fluctuating in-phase charge transfer induces the dynamic symmetric FS distortions illustrated in Fig. \ref{Fig4_FS_distortion}(i). The broad A$_{1g}$ peak at  350 cm$^{-1}$ can be interpreted as a collective excitation associated to a A$_{1g}$ type of dynamic FS deformation.

In summary, we used Raman scattering to investigate the SC gap and collective excitations in SC BaFe$_2$(As$_{0.5}$P$_{0.5}$)$_2$ ($T_c \approx 16$ K). We observe a SC pair breaking peak at 54 cm$^{-1}$ (6.7 meV) that appears below $T_c$ in the A$_{1g}$ channel. The linear decrease of the spectral weight suggests nodes in the SC gap structure. We identify spectral features associated with Pomeranchuk oscillations in the A$_{1g}$, B$_{1g}$ and B$_{2g}$ symmetry channels. Unexpectedly, these features are well-defined, which we attribute to the fact that these frequencies are larger than the Fermi energy of this system, thus preventing overdamping by electron-hole continuum. The degeneracy of the Pomeranchuk mode in the B$_{1g}$ and B$_{2g}$ symmetry channels is explained by the mixing of these two symmetry channels due to the removal of the $\sigma_v$ and $\sigma_d$ operations caused by As/P disorder. This disorder manifests itself by the absence of the pnictogen A$_{1g}$ phonon normally detected in other Fe-based superconductors. Interestingly, the temperature regime at which the Pomeranchuk oscillations get underdamped coincides with a crossover regime from NFL to FL determined from transport. Our results indicate that the Pomeranchuk instability plays a critical role in the low-energy physics of this Fe-based superconductor. 

This work was supported by grants from MOST (2015CB921301) and NSFC (11274362) of China. S.-F.W., H.-H.K., A.L., and G.B. acknowledge support from the NSF Grant No. DMR-1104884 for spectroscopic characterization of pnictide materials. H.-H.K. and G.B. acknowledge support from the US Department of Energy, Basic Energy Sciences, and Division of Materials Sciences and Engineering under Grant No. DE-SC0005463 for the data analyses.

\bibliography{biblio_long}

\end{document}